\documentclass[11pt,a4paper,DIV=10]{scrartcl}
\usepackage[utf8]{inputenc}
\usepackage{amsmath}
\usepackage{amsfonts}
\usepackage{amssymb}
\usepackage[dvipsnames]{xcolor}
\usepackage{graphics}
\usepackage[super,sort&compress,comma]{natbib} 
\usepackage[version=3]{mhchem}
\usepackage{textcomp}
\usepackage{mathpazo}
\usepackage{hyperref}
\usepackage{multirow}
\usepackage{graphicx}
\hypersetup{
    colorlinks = true,
    urlcolor = {Blue},
    citecolor = {Cerulean},
    linkcolor = {Bittersweet}}

\renewcommand{\d}{\mathrm{d}}

\title{Perovskite/silicon tandem solar cells: Effect of luminescent coupling and bifaciality}
\author{\large Klaus J\"{a}ger,\thanks{Corresponding author; e-mail: \href{mailto:klaus.jaeger@helmholtz-berlin.de}{\texttt{klaus.jaeger@helmholtz-berlin.de}}} \thanks{Helmholtz-Zentrum Berlin f\"ur Materialien und Energie, Albert-Einstein-Stra\ss e 16, D-12489 Berlin.} \thanks{Zuse Institute Berlin, Takustra\ss e 7, D-14195 Berlin.} ~Peter Tillmann,$^{\dag\ddag}$ Eugene A.\ Katz,\thanks{Dept. of Solar Energy and Environmental Physics, The Jacob Blaustein Institutes for Desert Research, Ben-Gurion University of the Negev, Sede Boqer Campus, 8499000 Israel.} ~and Christiane Becker$^\dag$}

\begin{document}
\maketitle

\begin{abstract}
    The power conversion efficiency of the market-dominating silicon photovoltaics approaches its theoretical limit. Bifacial solar operation with harvesting additional light impinging on the module back and the perovskite/ silicon tandem device architecture are among the most promising approaches for further increasing the energy yield from a limited area. Here, we calculate the energy output of perovskite/silicon tandem solar cells in monofacial and bifacial operation considering, for the first time, luminescent coupling between two sub-cells.  For energy yield calculations we study idealized solar cells at both, standard testing as well as realistic weather conditions in combination with a detailed illumination model for periodic solar panel arrays. Considering typical, experimental photoluminescent quantum yield values we find that more than 50\% of excess electron-hole pairs in the perovskite top cell can be utilized by the silicon bottom cell by means of luminescent coupling. As a result, luminescent coupling strongly relaxes the constraints on the top-cell bandgap in monolithic tandem devices. In combination with bifacial operation, the optimum perovskite bandgap shifts from 1.71\,eV to the range 1.60-1.65\,eV where already high-quality perovskite materials exist. The results can hence change a paradigm in developing the optimum perovskite material for tandem solar cells.
\end{abstract}


\section{Introduction}
\begin{figure}
 \centering
 \includegraphics{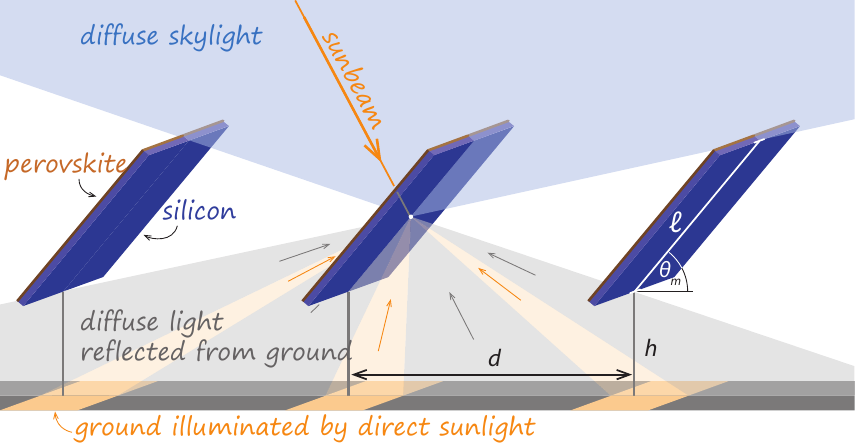}
 \caption{Illustrating the illumination components reaching a bifacial solar module in a large photovoltaic field: both, the front and back sides, can be illuminated by direct sunlight, diffuse skylight, and light from the ground, which can originating from direct sunlight or diffuse skylight. The photovoltaics field is characterized by the module length $\ell$, height of the modules above the ground $h$, module tilt angle $\theta_m$, distance between rows of modules $d$ and albedo of the ground $A$.\cite{jaeger:2020oe}}
\label{fig:bifacial-sketch}
\end{figure}

Monofacial silicon solar cells currently dominate the photovoltaic (PV) market.\cite{ITRPV2020} Their practical efficiencies meanwhile approach the theoretical limit of around 29.4\%,\cite{Richter2013ReassessmentCells} such that innovative technologies and concepts are required to increase the energy yield on limited areas. One approach is using bifacial solar systems that cannot only utilize light, which falls onto the front side of the PV module, but also light reaching the back side,\cite{kopecek_towards_2018, liang_review_2019} as illustrated in Fig.\ \ref{fig:bifacial-sketch}. Bifacial photovoltaic power plants demonstrated $>20\%$ enhanced annual energy yield in comparison to a monofacial power plant of a similar size.\cite{Ishikawa_Bifacial_2016} Modern silicon solar cell concepts with passivated emitter rear contact (PERx), heterojunction (SHJ) or integrated back contact (IBC) enable bifacial solar cell operation at low additional cost. Due to these reasons the International Technology Roadmap for Photovoltaics predicts nearly 70\% market share for bifacial solar cells in 2030.\cite{ITRPV2020} 

A second method to increase the energy output from a photovoltaic system on limited area is the multi-junction approach where multiple solar cells with different band gaps are stacked on top of each other. These different materials exhibit complementary electronic bandgaps such that the high energy photons of solar irradiation are absorbed by the high-bandgap materials on top, while the lower energy photons are absorbed by the lower bandgap material at the bottom. As a result the excess photon energy losses are reduced and conversion efficiencies increase, significantly overcoming the efficiency limit of silicon single-junction solar cells.

A currently widely investigated technology for large scale applications is the combination of silicon and perovskite solar cells in a tandem device.\cite{werner:2017} High efficiencies, a tunable bandgap, external photoluminescent quantum yields up to 10\% \cite{liu_open-circuit_2019} and low-cost fabrication processes make perovskites an attractive tandem partner for established silicon photovoltaics. The current record efficiencies for perovskite/silicon tandem solar cells are 29.15\% \cite{ashouri:2020} for monolithic two-terminal (2T) and 28.2\% \cite{chen:2020} for stacked four-terminal (4T) devices, respectively,  bearing the potential for power conversion efficiencies as high as $\approx44\%$\cite{leijtens_opportunities_2018} assuming radiative recombination the only recombination channel and standard test conditions (STC), i.e.\ 25\textdegree C temperature and 1000\,W/m$^2$ solar irradiance with AM1.5g spectral distribution.\cite{iec:60904-3} The monolithic tandem configuration has (among others) the advantage of requiring only two external contacts and one maximum power point tracker, enabling module related costs comparable to single-junction devices. \cite{jost:2020} Under STC, the theoretical power output of silicon-based monolithic tandem solar cells, however, reveals a sharp maximum at a top-cell bandgap around 1.71\,eV limiting the choice of available perovskite top cell materials. The reason for the sharp optimum is the current matching requirement in a monolithic series-connected tandem device, i.e. the top cell bandgap has to be tuned such that the numbers of generated electrons are the same for the the top cell and the bottom cell. However, perovskites with band gaps above 1.7\,eV often suffer from low electronic quality resulting in reduced solar-cell efficiencies.\cite{unger:2017}

In recent years, bifacial perovskite/silicon tandem solar cells were extensively investigated.\cite{schmager_methodology_2019, onno_predicted_2020, imran_high-performance_2018,dupre_design_2020, asadpour_bifacial_2015} In particular, Onno \emph{et al.}\ found that the range of appropriate top-cell bandgaps broadens in a bifacial tandem-cell configuration.\cite{onno_predicted_2020} This is in line with thermodynamic consideration by Khan \emph{et al.}\cite{khan2015} Additional photons absorbed in the silicon bottom cell from rear side illumination allow for a lower bandgap of the (perovskite) top cell at current-matching conditions. 

\begin{figure}
 \centering
 \includegraphics{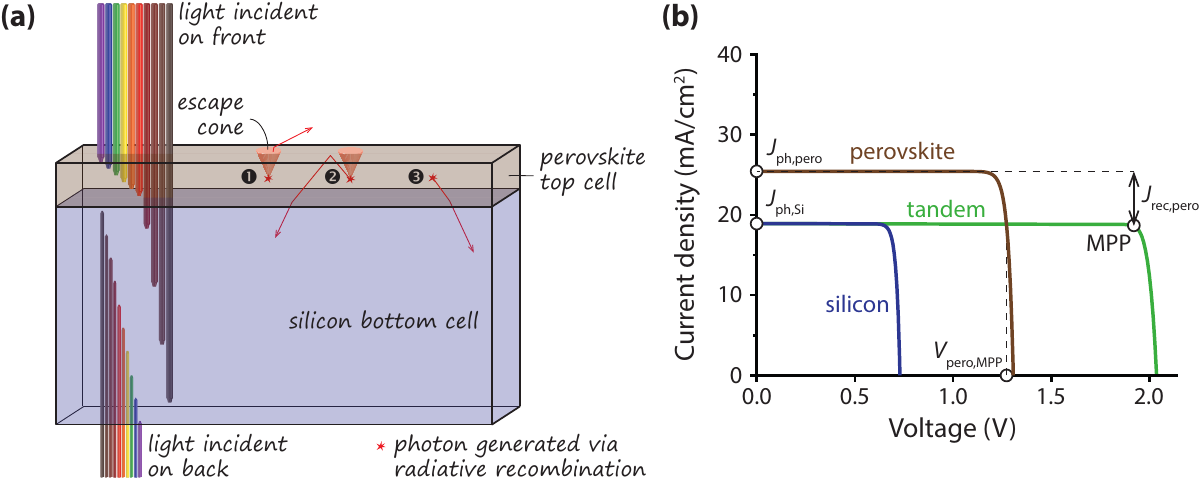}
 \caption{(a) Illustrating luminescent coupling (LC) in a perovskite/silicon tandem solar cell. A photon, which is generated in the perovskite top cell via radiative recombination can (1) either leave the perovskite cell if its direction is within the escape cone, or (2) it undergoes total internal reflection and is redirected downward such that it can enter the silicon cell, just as (3) a photon that is emitted into the lower hemisphere. More details can be found in appendix \ref{sec:lc}. (b) An example for $JV$ curves of a bottom-cell limited tandem cell (green) and the perovskite (brown) and silicon (blue) subcells illuminated under STC. Here, the perovskite is simulated with a bandgap of 1.6\,eV and generates a higher photocurrent density $J_\text{ph,pero}$ than the silicon subcell with  $J_\text{ph, Si}$. At the maximum power point (MPP) of the tandem cell significantly less current density is extracted from the perovskite cell than generated. The excess current density $J_\text{rec, pero}$ can be re-utilized via luminescent coupling to increase the photocurrent density of the silicon subcell.}
\label{fig:lc-sketch}
\end{figure}

One aspect of perovskite-based tandem photovoltaic operation has not been considered so far: luminescent (or radiative) coupling between the different subcells in the device, i.e.\ the re-absorption of luminescent photons emitted by the high-bandgap top cell in the low-bandgap bottom cell. This effect is well-known in multi-junction solar cells based on III-V semiconductors, where luminescent-coupling efficiencies
above 30\% were reported.\cite{walker:2015} Already in 2002, Brown and Green identified luminescent coupling as a means to reduce spectral mismatch in two-terminal tandem solar cells.\cite{brown_radiative_2002} While the effect of luminescent coupling is negligible at current-matching conditions, a considerable positive effect appears in non-current-matched, bottom-cell limited devices. \cite{steiner_non-linear_2012, shvarts:2013, chan_practical_2014, friedman:2014} Similar to bifacial cell operation luminescent coupling, i.e.\ the re-absorption of luminescent photons emitted by the high-bandgap cell in the low-bandgap cell, results in more photons absorbed in the silicon bottom cell, as illustrated in Fig.\ \ref{fig:lc-sketch}(a). To the best of our knowledge, luminescent coupling has not been investigated experimentally for perovskite-based multi-junction solar cells yet. 

In this study, we theoretically investigate how bifacial illumination and luminescent coupling affect the performance of perovskite/silicon tandem solar cells. We use idealized solar-cell models for these calculations: Shockley-Queisser's detailed balance limit\cite{Shockley1961DetailedCells} for the perovskite top cell and the Richter limit\cite{Richter2013ReassessmentCells} for the silicon bottom cell, which also incorporates Auger recombination. For the perovskite cell operation under one Sun, Auger recombination is negligible.\cite{wang:2018} Using these models, we first assess, how illumination from the back side and luminescent coupling affect the tandem-cell performance under standard test conditions. Then, we use optical simulations\cite{santbergen:2017} to estimate, how much of light from radiative recombination in the perovskite leaves the cell towards the Sun in a single-junction cell configuration and how much will reach the silicon subcell in a tandem stack. This allows us to relate measured \emph{external} quantum photoluminescence efficiency in a single-junction perovskite cell to the reasonable \emph{internal} quantum efficiency, and subsequently to evaluate, which range of luminescent-coupling efficiencies is realistic in tandem devices. Last, we estimate the energy yield using weather data from a climatic zone with high diffuse illumination ratio. For this we apply a detailed illumination model, which takes direct sunlight, diffuse skylight, shadowing by other modules and reflection from the ground into account.\cite{jaeger:2020oe} We finally discuss how all the realistic deviations from standard test conditions considered in this study -- (1) bifacial irradiation, (2) luminescent coupling and (3) weather conditions with high diffuse illumination ratio –- influence the constraints for the perovskite top cell bandgap.

\section{Modelling details}
\subsection{Electrical solar cell model}

To calculate the current density-voltage $(JV)$ characteristic of the PV modules, the irradiance values on the front and back sides are used as input for the electrical model. In this paper we use highly idealized solar cell models:

For the perovskite top cell we assume that all photons with energy higher that the cell band gap are absorbed and every absorbed photon generates one electron-hole pair. Hence, the maximum achievable photocurrent density is given by
\begin{equation}
    \label{eq:jsc-pero}
    J_\text{ph, pero} = e\int_0^{\lambda_\text{pero}} \Phi_f(\lambda)\,\text{d}\lambda,
\end{equation}
where $e$ is the elementary charge, $\Phi_\text{f}$ is the photon flux reaching the module at the front, and $\lambda_\text{pero}$ is the wavelength corresponding to the perovskite bandgap. In a monolithic tandem device this value is only achieved in case of a limiting top cell, i.e.\ less or equal photons absorbed in the perovskite than in the silicon. The $JV$ characteristic is calculated according to the \emph{Shockley-Queisser} (SQ) limit,\cite{Shockley1961DetailedCells} where only radiative recombination is considered. In the SQ limit, both external (ELQE) and internal (ILQE) luminescence quantum efficiencies are equal to 100\%. The former is the number of photons emitted into free space relative to the number of electron–hole pairs generated by light absorption in a solar cell. The latter is a ratio between the number of electron–hole pairs recombined radiatively to the entire number of the recombined pairs. The SQ limit is briefly summarized in appendix \ref{sec:pero-model}.

For the silicon bottom cell, the perovskite top-cell acts as a filter for the short wavelengths up to the perovskite bandgap. However, the perovskite cell also may emit light, which can be utilized by the bottom cell via \emph{luminescent coupling}, which is discussed below. Additionally, Auger recombination must be considered for a silicon cell. We implement this using an idealized model by Richter and coworkers;\cite{Richter2013ReassessmentCells} the details are given in appendix \ref{sec:si-model}. 

In a high-end solar cell made of a direct bandgap semiconductor a significant fraction of the absorbed photons, which are not extracted as electrical current, will be re-emitted as light via radiative recombination. An electrically independent solar cell operated at maximum power point only has a small recombination current because almost all charge carriers are extracted. However, in a two-terminal tandem cell, where the top and bottom cells are electrically connected in series, the same current density flows through bottom and top cell. If the generated photocurrent density and the extracted current density deviate strongly from each other, as illustrated in Fig.\ \ref{fig:lc-sketch}(b), significant recombination will be present in the top cell. If the recombination is radiative, the re-emitted light from the top cell can be absorbed and utilized by the bottom cell, which is known as \emph{luminescent coupling} (LC). In perovskite/silicon tandem solar cells we only need to consider light emitted by the perovskite cell, which can be absorbed by the silicon bottom. The silicon cell itself will hardly emit light because of the indirect bandgap of silicon. Further, the energy of the emitted photons would be close to the silicon bandgap and hence cannot be absorbed by perovskite with a larger bandgap than silicon. For the maximum achievable short-circuit current density in the Si bottom cell we find
\begin{equation}
    \label{eq:jsc-Si}
    \begin{aligned}
        J_\text{ph, Si}(V_\text{pero}) =& e\int_{\lambda_\text{pero}}^{\lambda_\text{Si}} A(\lambda)\Phi_f(\lambda)\,\text{d}\lambda + e\int_0^{\lambda_\text{Si}} A(\lambda)\Phi_b(\lambda)\,\text{d}\lambda \\
        &+ \eta_\text{LC} \left[J_\text{ph,pero}-J_\text{pero}(V_\text{pero})\right]
    \end{aligned}
\end{equation}
with the absorption in silicon $A(\lambda)$. We calculate the absorption in Si according to the Tiedje-Yablonovitch limit for a silicon wafer thickness of 300 $\mu$m as described in appendix \ref{sec:si-model}. $J_\text{pero}(V_\text{pero})$ is the current density at the working point of the perovskite cell. The term $\left[J_\text{ph,pero}-J_\text{pero}(V_\text{pero})\right]$ corresponds to excess electron-hole pairs generated in the perovskite top cell, which cannot be extracted from the monolithic tandem device, e.g. due to a limiting bottom cell. These excess electron-hole pairs can recombine radiatively and be re-absorbed by the silicon with $\eta_\text{LC}$ being the efficiency of this luminescent coupling. Here we also accounted for light that hits the solar cell at the back, $\Phi_b$. For monofacial cells we have $\Phi_b\equiv 0$. Further, $\lambda_\text{Si}$ is the wavelength corresponding to the silicon bandgap. More details about luminescent coupling are given in appendix \ref{sec:lc}. 

Since we assume zero series resistance and infinitely large shunt resistance of the cells, for both subcells the electric current density $J$ can be directly calculated from the photocurrent density $J_\text{ph}$ and the voltage-dependent recombination current density $J_\text{rec}$,
\begin{equation}
    J = J_\text{ph}- J_\text{rec}(V),
\end{equation}
where details about $J_\text{rec}$ for the perovskite and silicon subcells are given in appendix \ref{sec:pero-model} and appendix \ref{sec:si-model}, respectively.

For two-terminal cells, where the same current density flows through both cells, we have
\begin{equation}
    J_\text{cell} = J_\text{ph,Si}- J_\text{rec,Si}(V_\text{Si}) = J_\text{ph,pero}- J_\text{rec,pero}(V_\text{pero}).
\end{equation}
We calculate the $JV$ characteristic of the tandem solar cell by numerically inverting $J_\text{rec,Si}(V_\text{Si})$ and $J_\text{rec,pero}(V_\text{pero})$ such that we have functions of $J_\text{rec,pero}$ and $J_\text{rec,Si}$, respectively.  From the $JV$ curve the output power density of the cell can be directly calculated as
\begin{equation}
    \begin{aligned}
        P_\text{cell} &= J_\text{cell}\left[V_\text{Si}(J_\text{rec,Si}) + V_\text{pero}(J_\text{rec,pero})\right],\\
        P_\text{mpp} &= \max_{J_\text{cell}} \left[ P_\text{cell}\right].
    \end{aligned}
\end{equation}

Tandem solar cells can also built in four-terminal configuration, where the two subcells are electrically independent and can operate at their individual maximum power points,
\begin{equation}
    P_\text{mpp} = \max_{J_\text{Si}}\left[J_\text{Si} \cdot V_\text{Si}(J_\text{rec,Si})\right] + \max_{J_\text{pero}}\left[J_\text{pero} \cdot V_\text{pero}(J_\text{rec,pero})\right].
\end{equation}

\subsection{Optical model}

In order to estimate the effect of luminescent coupling in realistic perovskite-tandem solar cells, we apply optical modelling. In this paper, we use the \texttt{MATLAB}-based tool \texttt{GenPro4}, which can calculate the absorption profile in solar-cell structures using the net radiation method.\cite{santbergen:2017} This tool treats light coherently in thin layers but incoherently in thick layers. Because \texttt{GenPro4} only can treat light that falls onto a layer stack from the exterior, we split the simulations in two: one simulation treating the layer stack above the perovskite layer, the other layer stack treating the layers below. Details on these calculations are given in appendix \ref{sec:opt}.

\subsection{Energy yield calculation}
We calculate the overall energy yield for different scenarios using a simulation approach that combines several sub models. For calculating the spectral irradiance at the front and back sides of a solar module in a big PV field, we employ a recently developed illumination model.\cite{tillmann:2020, jaeger:2020oe} The PV field is considered so large that boundary effects can be neglected. As schematically illustrated in Fig.\ \ref{fig:bifacial-sketch}, the illumination model considers four components reaching the module front: direct sunlight, diffuse skylight, diffuse light from the ground, which originates from direct sunlight reaching the ground and diffuse skylight reaching the ground. Further, the same four components must be considered reaching the back-side of the module. Hence the illumination model considers eight components in total.

The illumination model uses the following input parameters: first, the geometrical parameters of the PV field, which are sketched in Fig.\ \ref{fig:bifacial-sketch}: module length $\ell$, mounting height $h$, module spacing $d$ and tilt angle $\theta_m$. Secondly, the albedo (i.e. the reflectivity) of the ground, which is highly dependent on the material properties of the ground. While grass typically exhibits albedo values around 20\%, grey and white gravel have albedo values of 30\% and 50\%, respectively, and snow reaches albedo values up 70\% \cite{taiyang:2018}. In this work, we assume the albedo to be independent of the wavelength with $A=30\%$, which is a rather conservative estimate with realistic room for improvement.  Thirdly, the (spectral) \emph{direct normal incidence} (DNI) and the \emph{diffuse horizontal incidence} (DHI) for different instants of time. We retrieve these data from the National Solar Radiation Data Base (NSRDB) operated by NREL.\cite{wilcox:2008} They publish hourly spectral direct and diffuse irradiance for a \emph{typical meteorological year} (TMY).

With the spectral irradiance on the front and back sides we can calculate the generated photocurrent densities in the top and bottom cells using eqn (\ref{eq:jsc-pero}) and (\ref{eq:jsc-Si}). We calculate the full $JV$-characteristics for every hour in the TMY data set and take the appropriate maximum to get the maximum power output of the cell according to eqn (\ref{eq:power-num}) and (\ref{eq:power-num2}). By integrating over all hourly data points in the data set for one year we obtain the annual energy yield.

\section{Results and discussion}

\subsection{Tandem-cell operation under standard testing conditions}

\begin{figure*}
 \centering
 \includegraphics[width=\textwidth]{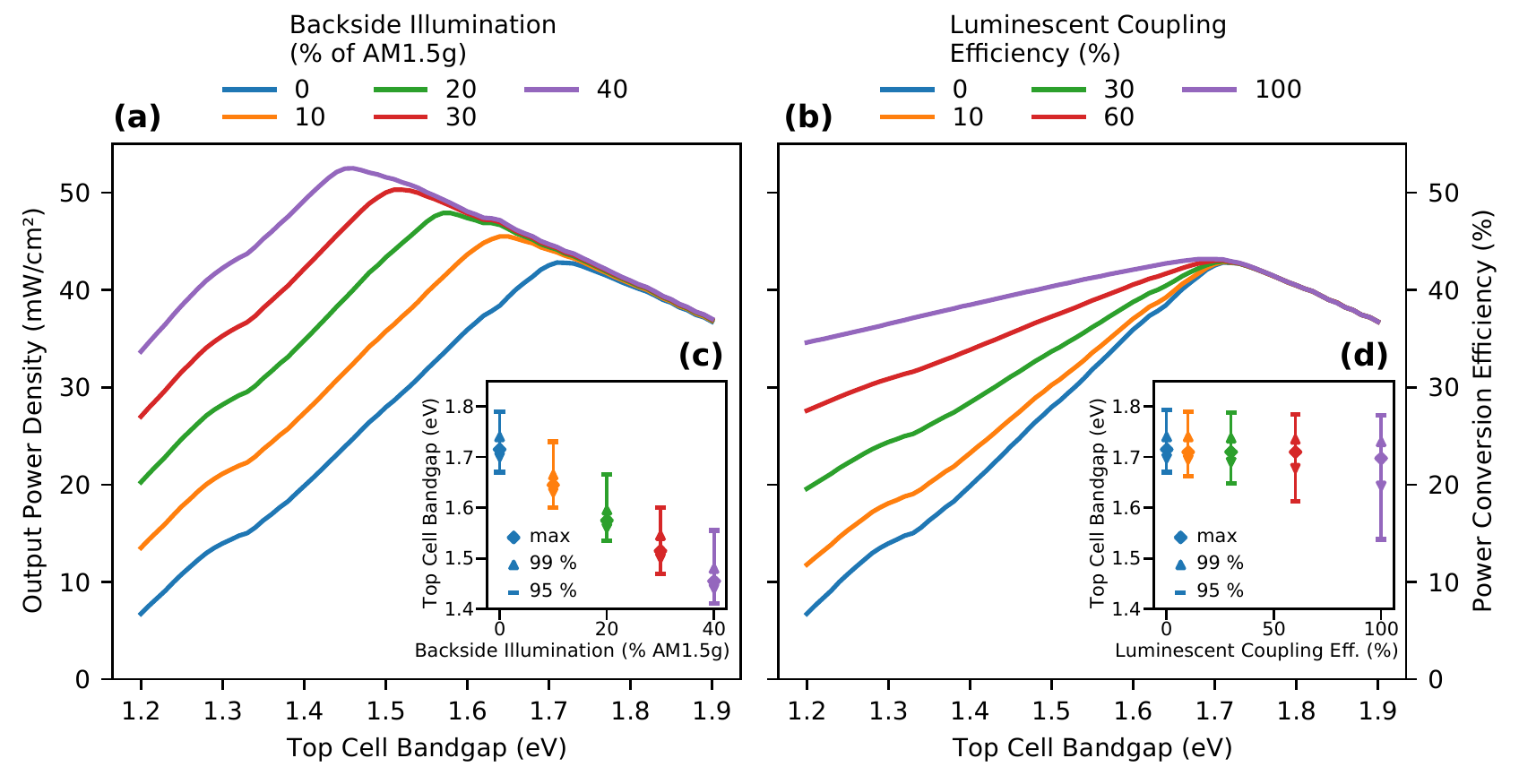}
 \caption{Maximum output power density of two-terminal tandem solar cells as function of the top-cell bandgap for different levels of  (a) backside illumination and (b) luminescent coupling efficiencies under standard test conditions. The insets show the optimal top-cell bandgap for different levels of (c) backside illumination and (d) luminescent coupling efficiencies under standard test conditions. The diamonds mark the ideal bandgap with maximum power output; the arrowheads and the dash marks span the ranges where at least 99\% and 95\% of the maximum output power density are achieved. Note: For the graph with varying backside illumination no luminescent coupling is assumed and for varying luminescent coupling efficiencies no backside illumination is present. The bottom cell bandgap is 1.12\,eV in all cases.}
\label{fig:eff_tandem_stc}
\end{figure*}

Figure \ref{fig:eff_tandem_stc} shows the effect of the top cell bandgap on the maximum output power density of a two-terminal tandem solar cell for various levels of backside illumination (a) and luminescent coupling (b) under standard testing conditions. Without either backside illumination or luminescent coupling the optimal bandgap of the perovskite cell for maximum power output density is 1.71\,eV, where the same current densities are generated in the top and bottom cells. For other top-cell bandgaps, the generated current densities differ from each other. Only the lower current density can flow through the solar cell, while the \emph{excess} current density is lost, which reduces the overall power conversion efficiency of the tandem solar cell. For a silicon-based tandem solar cell, the bandgap of the top cell absorber is critical to achieve current matching between the subcells. For a top-cell bandgap higher than the optimum, the current density generated in the top cell is below that generated in the bottom cell, the tandem cell is said to be ``top-cell limited''. For a top-cell bandgap lower than the optimum, the bottom-cell current density is lower; the cell is ``bottom-cell limited''.

With higher levels of backside illumination, as shown in Fig.\ \ref{fig:eff_tandem_stc}(a), the maximum power output density increases and the optimum top-cell bandgap shifts towards lower bandgaps. The backside illumination is exclusively absorbed in the bottom cell and cannot reach the top cell, leading to more generated electron-hole pairs in the bottom-cell. To match the photocurrent densities between the two subcells, the top-cell bandgap needs to be lowered, such that it can absorb more light. For top-cell bandgaps larger than 1.71\,eV, increased back-side illumination hardly affects the overall output power density, because here the tandem device is top-cell limited and the additional photocurrent generated in the bottom cell cannot be utilized.

As shown in Fig.\ \ref{fig:eff_tandem_stc}(b), increasing the luminescent coupling efficiency does not shift the position and height of the maximum output power density; however, the power output is increased for bandgaps below the optimum. For top-cell bandgaps above the optimum, luminescent coupling does not affect the performance, because here the cells are top-cell limited and the excess current in the bottom cell cannot be utilized for luminescent coupling.

The insets in Fig. \ref{fig:eff_tandem_stc} summarize these results. For a given scenario of backside illumination or luminescent coupling the optimal bandgap and the range of 99\% and 95\% of the maximum output power density are shown. With increasing backside illumination, the optimal top-cell bandgap shifts to lower values, while sensitivity is unchanged. For luminescent coupling, the optimal bandgap remains unchanged but the 99\%- and 95\% bands broaden towards lower bandgaps.

\subsection{Estimating reasonable values of luminescent-coupling efficiency}

Now, as we have studied how luminescent coupling can improve the performance of bottom-cell limited tandem solar cells [see Figs.\ \ref{fig:eff_tandem_stc}(b,d)], we investigate, which luminescent coupling efficiencies are realistic in perovskite/silicon tandem solar cells from an optical point of view. 

Increasing the power conversion efficiency of solar cells towards the theoretical limit can be realized by improving the external luminescence quantum efficiency (ELQE) of the cell in open circuit (OC), or in the other words –- by suppressing non-radiative recombination.\cite{Rau2007ReciprocityCells, Miller2012StrongLimit} Despite the direct bandgap of metal halide perovskite semiconductors, initially reported ELQE values for perovskite solar cells were extremely low ($\approx10^{-4}\%$).\cite{Tvingstedt2014RadiativeCells} Then, tremendous growth was demonstrated for perovskite solar cells reaching an ELQE of 0.5\%,\cite{Bi2016EfficientPerovskites} which is equal to the record for silicon cells.\cite{Green2012RadiativeCells} Recently, Liu and co-workers realized a single junction perovskite solar cell with 8.4\% ELQE.\cite{liu_open-circuit_2019} Note that record ELQE values of the champion GaAs cells do not exceed 25\%,\cite{Kayes201127.6Illumination, Braun2013PhotovoltaicEmission} even though internal luminescence quantum efficiency (ILQE) values of 99.7\% have experimentally been shown for GaAs devices.\cite{Schnitzer1993UltrahighHeterostructures}

\begin{figure}
 \centering
 \includegraphics{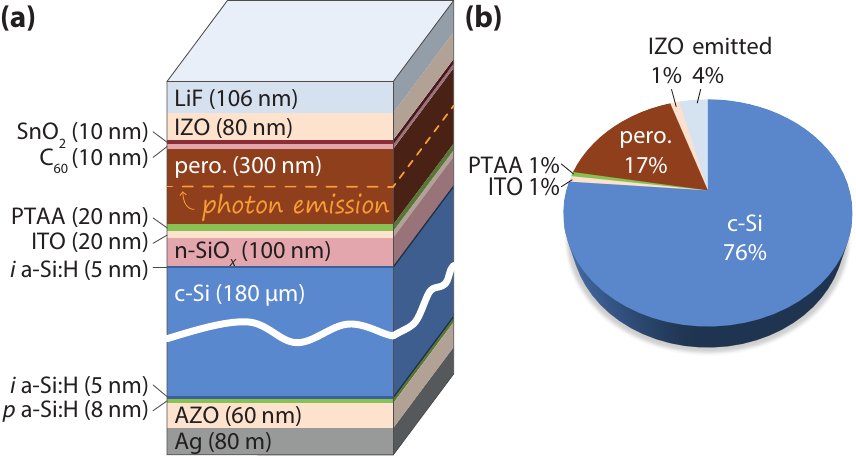}
 \caption{\label{fig:lc-tandem-optical}(a) The tandem solar cell structure used for estimating the fraction of photons, which are generated in the perovskite layer and reach the silicon wafer. The structure is based on recent high-end perovskite/silicon tandem solar cells.\cite{jost:2018, koehnen:2019} The dotted line indicates the middle of the perovskite layer (150\,nm depths). In our calculations, we assumed the light emission from this depth. (b) Relative distribution of photons with 795\, nm wavelength, which are isotropically emitted in the center of the perovskite layer. While around 76\% are absorbed by the silicon wafer, around 17\% are reabsorbed by the emitting perovskite layer. Only $\approx$4\% leave the solar cell structure.}

\end{figure}

As a first step to estimate the luminescent coupling efficiency for a cell with the experimentally measured 8.4\% ELQE, we calculate the fraction $E_t^\text{int}$ of light generated in the perovskite layer, which leaves the solar cell structure, using the optical simulation tool \texttt{GenPro4}. We assume a perovskite thickness of 400\,nm and an emission wavelength of 795\,nm, which corresponds to the bandgap of the perovskite methylammonium lead iodide (\ce{MAPbI3}) of 1.56\,eV, in accordance with the device architecture used by Liu \emph{et al.}\cite{liu_open-circuit_2019} As shown in appendix \ref{sec:opt}, we revealed $E_t^\text{int}=7.8\%$ for this configuration, which is independent of the emission depths in the perovskite layer. The rest of the generated light cannot leave the solar cell structure, because it either radiates in directions outside the emission cone, which has an opening angle of 23.8\textdegree\ for \ce{MAPbI3} at 795\,nm,\cite{guerra:2017} or it is absorbed before it can leave the solar cell. The experimental ELQE (8.4\%) being larger than the numerical value $E_t^\text{int}=7.8\%$ shows that a high ILQE was achieved. For semiconductors with high ILQE, photon recycling,\cite{Brenes2019BenefitCells} i.e.\ the re-absorption of previously emitted photons within the perovskite, can increase the ELQE to values higher than what would be expected from the optical simulations without photon recycling.\cite{Cho2020TheDiodes} An experimental proof of internal photon recycling in perovskite solar cells was given by Pazos-Out\'{o}n \emph{et al.}\cite{pazosouton:2016} Further, Braly and coworkers demonstrated perovskite films with 90\% internal photoluminescence quantum efficiency.\cite{braly:2018}

We can estimate the ILQE using a simple model for a cell in open circuit condition where the charge carriers created by the external light source can undergo a chain of emission and reabsorption events. In a first step the charge carriers can be either recombine radiatively with probability ILQE or non-radiatively with probability $\left(1-\text{ILQE}\right)$. In the next step the emitted photons can either leave the cell with probability $E_t^\text{int}$, be absorbed parasitically in non-active areas with probability $A_\text{para}$ or reabsorbed in the perovskite with probability $A_\text{pero} = 1-E_t^\text{int}-A_\text{para}$. $A_\text{pero}$, $E_t^\text{int}$ and $A_\text{para}$ can be extracted from the optical simulations described in Appendix \ref{sec:opt}. The reabsorbed light can undergo the same processes as the directly absorbed light from an external light source. This chain of events can be represented as a geometric series to calculate the ELQE,
\begin{equation}
    \label{eq:pr}
    \begin{aligned}
        \text{ELQE} &= E_t^\text{int} \cdot \text{ILQE}\left[1 + A_\text{pero}\cdot \text{ILQE} + \left(A_\text{pero}\cdot \text{ILQE}\right)^2 + \ldots\right] \\
        &= \frac{E_t^\text{int} \cdot \text{ILQE}}{1-A_\text{pero}\cdot \text{ILQE}}.
    \end{aligned}
\end{equation}
This function can be inverted to retrieve ILQE,
\begin{equation}
    \label{eq:ilqe}
    \text{ILQE} = \left(\frac{E_t^\text{int}}{\text{ELQE}}+A_\text{pero}\right)^{-1}.
\end{equation}
Using eqn.\ (\ref{eq:ilqe}), we estimate the ILQE of the best cell from Liu \emph{et al.} \cite{liu_open-circuit_2019} to be around 65\%. This is in line with simulations from Cho \emph{et al.} on perovskite-based light emitting diodes, where they calculate that an ILQE of 60\% is sufficient to reach an ELQE equal to the purely optical expectation if photon recycling is considered.\cite{Cho2020TheDiodes}

\begin{figure*}
 \centering
 \includegraphics[width=\textwidth]{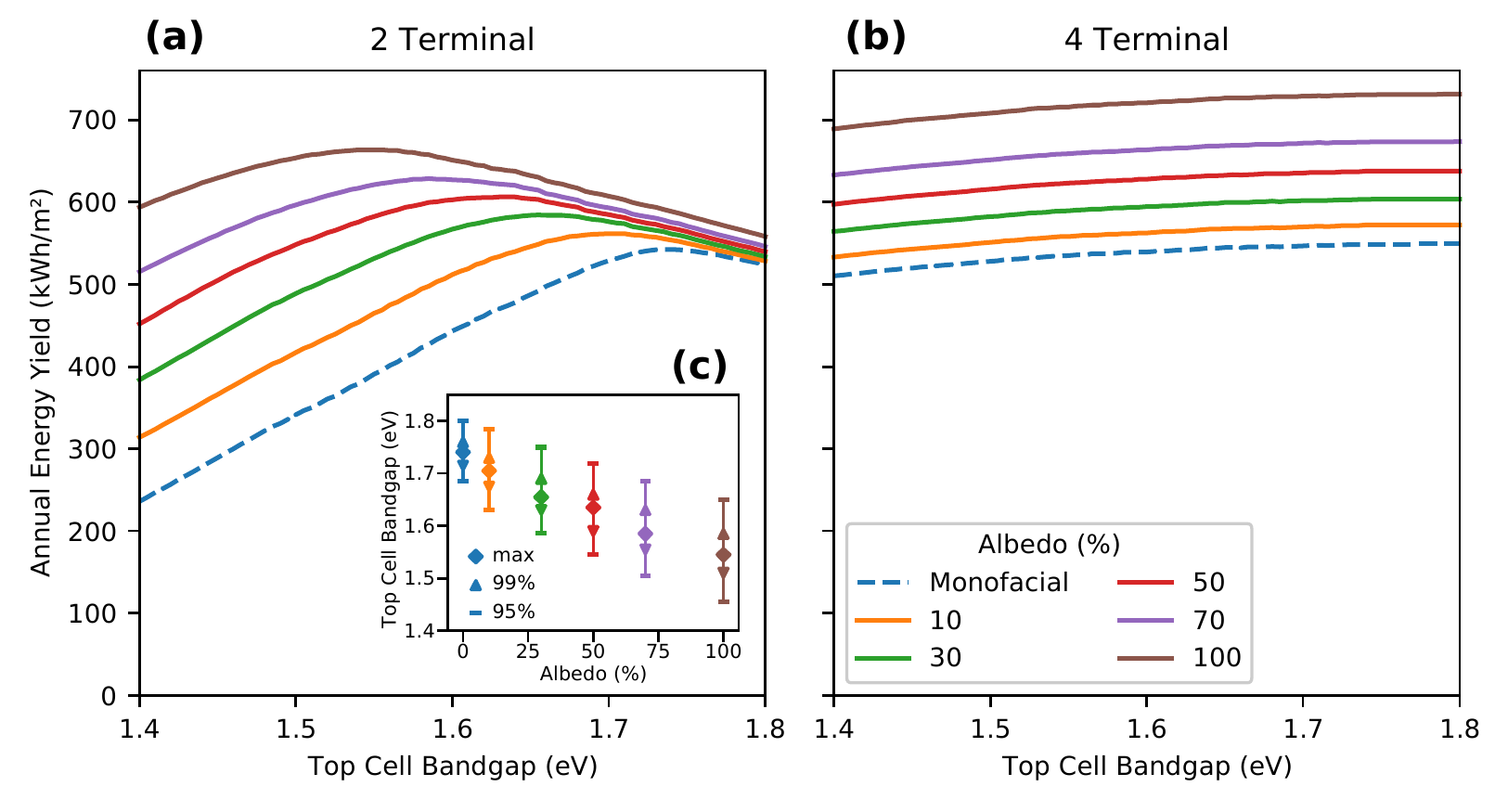}
 \caption{Energy yield for bifacial and monofacial tandem power plants simulated for Seattle with (a) two-terminal and (b) four-terminal cells connection for different albedo values. The inset (c) shows the optimal top-cell bandgap for different levels of albedo. The diamonds mark the ideal bandgap with maximum energy yield; the arrowheads and the dash marks span the ranges where at least 99\% and 95\% of the maximum energy yield is achieved.
 All simulations were performed with a module distance $d = 8$\,m and mounting height of $h=0.5$\,m. The tilt angle was optimized for every data point. Monofacial tandems are simulated with albedo $A=0\%$}
\label{fig:ey_no_lc}
\end{figure*}

Figure \ref{fig:lc-tandem-optical}(a) shows the perovskite/silicon tandem solar cell structure, which we used to study coupling of emitted light by the perovskite layer into silicon. This structure is based on recent high-end tandem solar cells,\cite{jost:2018, koehnen:2019} but in contrast to them we used \ce{MAPbI3} as perovskite material in order to be consistent with the single-junction results discussed above. For an emission wavelength of 795\,nm, 76\% of the light generated in perovskite reaches the silicon layer. This value is almost independent from the emission depths in the perovskite layer, as shown in appendix \ref{sec:opt}. Only 4\% of the generated light leave the solar cell structure into air and 17\% are reabsorbed in the perovskite layer, which can contribute to photon recycling. More details of the optical tandem-cell simulations are shown in Fig.\ \ref{fig:lc-tandem-optical-detail} in appendix \ref{sec:opt}.

We can estimate an upper bound for the luminescence coupling efficiency $\eta_\text{LC}^\text{max}$ by replacing $E_t^\text{int}$ with $A_\text{Si}$ in eqn (\ref{eq:pr}),
\begin{equation}
    \label{eq:lc-max}
        \eta_\text{LC}^\text{max} = \frac{A_\text{Si} \cdot \text{ILQE}}{1-A_\text{pero}\cdot \text{ILQE}}.
\end{equation}
For estimating $\eta_\text{LC}^\text{max}$ we use the values for 150\, nm emission depths, shown in Fig.\ \ref{fig:lc-tandem-optical-detail}(b): $A_\text{Si} = 0.763$ and $A_\text{pero} = 0.171$. Assuming $\text{ILQE} = 65\%$, just as for the single-junction cell discussed above, we find $\eta_\text{LC}^\text{max}\approx56\%$. However, it should be noted that Liu \emph{et al.}\ measured the ELQE  with an illumination of one sun without charge-carrier extraction (open circuit condition, in which all photo-generated carriers should recombine). When charge carriers are extracted in solar cell operation the ratio of radiative to non-radiative recombination might change considerably.\cite{Jia2015Bias-dependenceCells} Further research is needed to asses realistic radiative efficiencies at low recombination currents. In any case, we provide a positive answer on the fundamental question: a significant fraction of light emitted by the perovskite sub-cell can reach the silicon wafer. This can change a paradigm in developing optimal perovskite materials for efficient tandem solar cells.

\subsection{Energy yield under realistic weather conditions}

Under realistic conditions, the illumination on a solar module in a large photovoltaic field consisting of periodic rows of solar panels will significantly differ from standard testing conditions. The spectral distribution and irradiance of light in the outdoors is constantly changing and the illumination on the backside is highly dependent on the layout of the PV field. Fig \ref{fig:ey_no_lc} shows the result of energy yield calculations for bifacial and monofacial tandem solar modules for different bandgaps and varying levels of albedo in Seattle, USA and compares the performance of two- and four-terminal solar cells.
The four-terminal cells show only a small dependence on the top-cell bandgap with the optimum at the upper limit of the simulated range (1.8\,eV) and a monotonic decrease towards 1.5\,eV. Increasing the albedo increases the energy yield but leaves character of the bandgap dependence unchanged.

In contrast, the two-terminal cells are strongly affected by changing the top-cell band gap. Similar to the results for STC [Fig.\ \ref{fig:eff_tandem_stc}(a)], there is a well-defined maximum for the bandgap with reduced energy yield for higher or lower values. The ideal top cell bandgap for monofacial cells shifts from of 1.71\,eV for STC to 1.74\,eV for Seattle.

With increasing albedo, the optimal top-cell bandgap shifts to lower values. The additional light impinging onto the backside is exclusively absorbed by the bottom cell. Reducing the bandgap of the top cell will increase their photocurrent density at the cost of the bottom cell. Thus, the two subcells can be made current-matched again by reducing the top-cell bandgap.

\begin{table}
\centering
	\caption{\label{table:ey-2t}
	Results from energy yield calculations of two-terminal tandem cells for different albedo scenarios using average meteorological year data for Seattle with module height $h = 0.5$~m and module distance $d=8$~m. *``Bifacial Gain'' denotes the gain in irradiance.}
	\small
\begin{tabular}{lccccc}
       Type &  Albedo &  Bifacial Gain* & Opt.\ Bandgap & Energy Yield\\
            &     (\%) &       (\%) &     (eV) &        (kWh/m$^2$/a) \\\hline
 Monofacial &       \phantom{00}0 &      --- &     1.74 &        543 \\\cline{0-0}
   \multirow{5}{*}{Bifacial} &      \phantom{0}10 &      \phantom{0}5.5 &     1.70 &        562 \\
 &     \phantom{0}30 &     12.7 &     1.66 &        584 \\
 &     \phantom{0}50 &     19.7 &     1.64 &        606 \\
 &     \phantom{0}70 &     27.1 &     1.59 &        628 \\
 &     100 &     37.4 &     1.54 &        664 \\\hline
\end{tabular}
\end{table}

Table \ref{table:ey-2t} summarizes the results from the energy yield calculations for photovoltaic modules with two-terminal tandem cells for different albedo values. For a realistic albedo of $A = 30\%$ corresponding to grey cement \cite{Levinson:2002} the optimal bandgap shows a shift of $0.08$\,eV with respect to a monofacial cell. In this scenario the energy yield is increased by 7.5\%, which is significantly smaller then the 12.7\,\% gain of irradiance. 

One reason for the increase of energy yield being smaller than the increase of irradiance is that light reaching the back side can only be utilized with the single junction power conversion efficiency of the bottom cell. Further, for two-terminal tandem solar cells decreasing the top-cell bandgap to ensure current matching reduces the overall open-circuit voltage and hence the power conversion efficiency.

However, considering the electronic material quality of state-of-the-art perovskites,\cite{unger:2017} the effect of bandgap-shift might be relevant. While in principle organic/inorganic perovskites can be fabricated with continuously tunable bandgaps,\cite{Eperon2014FormamidiniumCells, unger:2017} not all bandgap-materials can be fabricated with the same electronic quality. Fabricating high-quality perovskite semiconductors with bandgaps in the range of 1.70-1.75\,eV is still a very challenging task and previous results show higher quality semiconductors in the region of 1.60-1.65 eV.\cite{jost:2020}

Operation of perovskite/silicon tandem solar cells in bifacial configuration allows to utilize 1.60-1.65 eV bandgap perovskites for optimal performance. This enables using current high-quality perovskite absorber layers in the tandem device.

\begin{figure}
 \centering
 \includegraphics{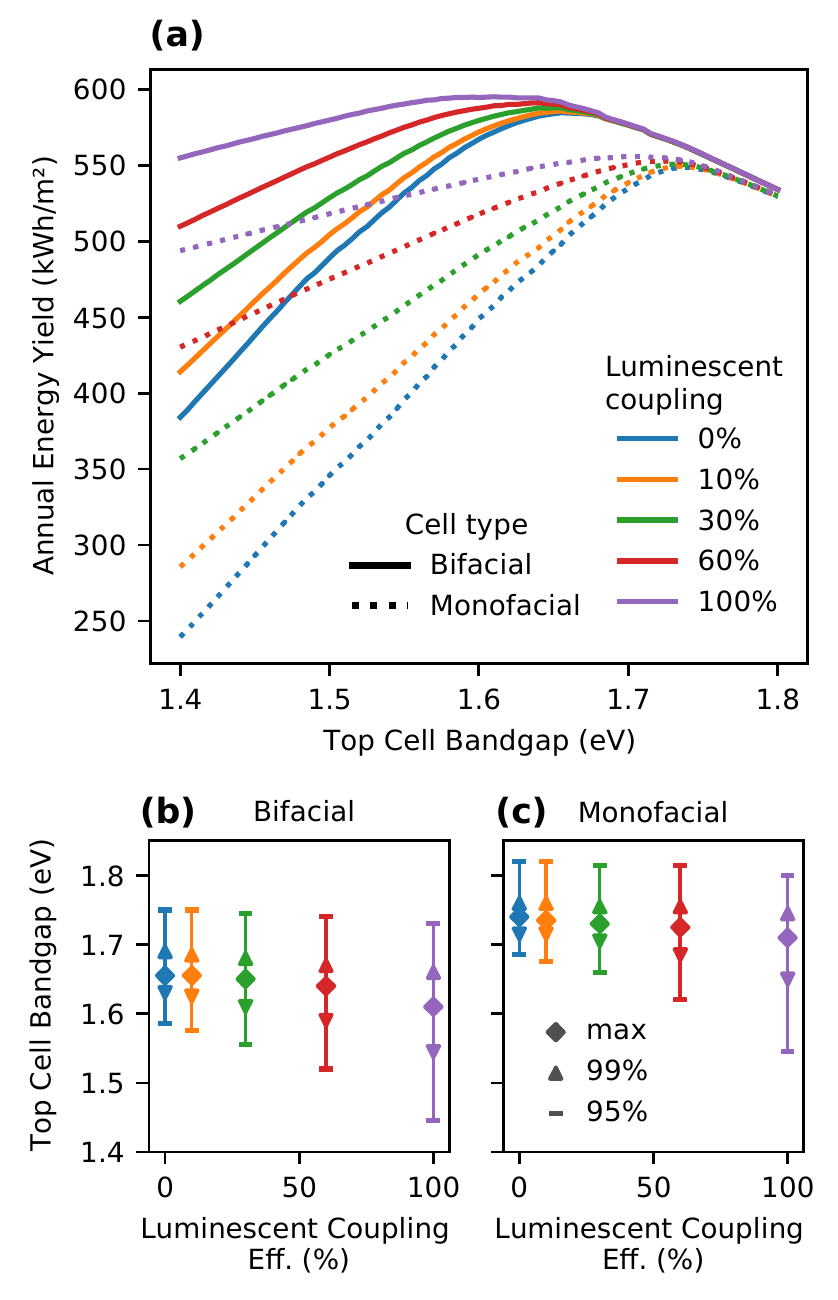}
 \caption{(a) Annual energy yield for mono- and bifacial two-terminal perovskite/silicon tandem solar cell modules simulated for Seattle with various levels of luminescent coupling. The sub figures shows the optimal top-cell bandgap for different levels of luminescent coupling of (a) bifacial and (c) monofacial tandem cells. The diamonds mark the ideal bandgap with maximum energy yield; the arrowheads and the dash marks span the ranges where at least 99\% and 95\% of the maximum energy yield is achieved  All simulations were performed with a module distance $d = 8$\,m and $d=0.5$\,m mounting height. Bifacial operation is calculated with albedo $A=30\,\%$. The module tilt angle $\theta_m$ was optimized for every data point.}
\label{fig:2t_lc_seattle}
\end{figure}

Figure \ref{fig:2t_lc_seattle} shows the effect of the top cell bandgap on the annual energy yield for mono- and bifacial two-terminal tandem PV modules simulated for Seattle, USA, with various levels of luminescent coupling. With an increasing luminescent coupling efficiency the energy yield becomes more and more independent from the bandgap of the top cell. Also, the maximum energy yield increases slightly and shifts a bit towards lower bandgaps. As the spectral distribution of outdoor illumination changes with time, there will always be situations where the top or bottom cells generate different photocurrent densities. Therefore, the optimal top-cell bandgap for outdoor performance will always be a compromise, which delivers the best balance over time.\cite{Horantner2017PredictingConditions} With increasing luminescent coupling efficiency the losses from periods, where the cell is bottom-cell limited, will become smaller while losses from top cell limitation are not affected.\cite{Okada2017AnalysisEffect} This explains the shift of the optimal bandgap to lower values, where the overall absorption in the top cell is increased. As an example, the energy yield of perovskite/silicon tandem solar cells with 1.64\,eV bandgap triple cation perovskite top cell is found to increase by 21.5\% when additionally considering a luminescent coupling efficiency of 30\% and bifacial operation on a 30\% reflective ground.

Four-terminal tandem solar cells barely show any performance improvement because of luminescent coupling, as both subcells are operated individually at their maximum power point, where only very little radiative recombination is present.

\section{Conclusions}
In conclusion, we calculated the energy yield of perovskite/silicon tandem solar cells considering luminescent coupling between the two sub-cells and bifacial illumination of the device. To do so, we first studied idealized solar cells by using the Shockley-Queisser limit and Richter's limit for the perovskite and the silicon sub-cells, respectively. We found that additional backside illumination around 10\%-20\% is sufficient to shift the optimum perovskite top-cell bandgap in two-terminal tandem solar cells from 1.71\,eV to the 1.60-1.64\,eV range. We further found that luminescent coupling can strongly reduce the current-mismatch if the tandem solar cell is bottom-cell limited.

As a second step, we performed optical simulations in order to evaluate the relevance of luminescent coupling for perovskite/silicon tandem solar cells. On the basis of experimental photoluminescent quantum yield values we found that more than 50\% of excess electron-hole pairs generated in the perovskite top cell can be re-used by the silicon bottom cell. Particularly for configurations with perovskite top cell bandgaps below the current matching optimum this significantly enhances the energy yield.

Last, we performed energy yield calculations based on typical meteorological year (TMY3) weather data of Seattle, USA, and applied an illumination model considering the spectral irradiance at the front and back sides of a solar module in a big PV field. In agreement with the calculations using standard testing conditions, we found that the operation of perovskite/silicon tandem solar cells in bifacial configuration allows to utilize 1.60-1.65 eV bandgap perovskites for optimal performance and luminescent coupling further minimizes the impact of current-mismatch in case of (silicon) bottom-cell limited devices, i.e. less photons absorbed in the silicon than in the perovskite absorber layer. The results can change a paradigm in developing the optimum perovskite material for tandem solar cells.

\appendix

\section{Perovskite solar cell}
\label{sec:pero-model}

We assume the current density ($J$) -- voltage ($V$) characteristic of the perovskite top cell to be that of a one-diode equation,
\begin{equation}
    J_\text{top}(V) = J_\text{0,top}\left[\exp\left(\frac{eV}{kT}\right)-1\right]-J_\text{ph,top},
\end{equation}
with the elementary charge $e$, the Boltzmann constant $k$, the temperature $T$ and the photon current density $J_\text{ph,top}$. The dark current density $J_\text{0,top}$ is calculated according to the Shockley-Queisser limit,\cite{Shockley1961DetailedCells} where only radiative recombination is considered in the solar cell, while non-radiative recombination processes like Shockley-Read-Hall recombination or Auger recombination are not accounted for. For calculating the radiative recombination rate, the solar cell is considered to be in thermal equilibrium with the surrounding at $T=300$~K and emits like a black body for photon energies higher or equal to the perovskite bandgap $E_\text{pero}$. Therefore, we find
\begin{equation}
    J_\text{0,top} = 2e\pi\int_{E_\text{pero}}^\infty{ \frac{2}{c^2h^3}\frac{E^2\,\d E}{\exp\left(\frac{E}{kT}-1\right)}},  
\end{equation}
where the first factor 2 arises from the solar cell emitting thermal radiation from both the front and back sides.
For the photon current density we assume that the perovskite absorbs step-like: all photons with energies larger or equal to the bandgap are assumed to be absorbed in the solar cell,
\begin{equation}
    J_\text{ph,top} = e\int_0^\frac{hc}{E_\text{pero}} \Phi_f(\lambda)\,\d \lambda,
\end{equation}
where $\Phi_f$ is the photon flux impinging on the solar cell front. (For calculating the SQ limit, $\Phi_f$ would be according to the AM1.5 standardized solar spectrum.)

\section{Silicon solar cell}
\label{sec:si-model}

Silicon is an indirect-bandgap material, which means that also Auger recombination has to be considered -- the Shockley-Queisser limit would overestimate the theoretical limit. Here, we follow a recent approach by Richter \emph{et al.}, who calculated the theoretical limit for silicon solar cells to be 29.4\%.\cite{Richter2013ReassessmentCells}

In the Richter limit, the $J$-$V$ characteristic of a silicon solar cell is given by
\begin{equation}
    J_\text{bot}(V) = e\cdot d_\text{Si}\cdot R_\text{intr}(V) - J_\text{ph,bot}
\end{equation}
with the silicon thickness $d_\text{Si}$ and the voltage-dependent intrinsic radiation rate $R_\text{intr}$, which accounts for both radiative and Auger recombination. We use a linear model for low doping concentrations according to eqn (21) of a work by Richter \emph{et al.} \cite{richter:2012}
\begin{equation}
    R_\text{intr} = np\left(8.7\cdot10^{-29}n_0^{0.91}+6.0\cdot10^{-30}p_0^{0.94}+3.0\cdot10^{-29}\Delta n^{0.92}+B_\text{tot}\right)
\end{equation}
with the electron and hole concentrations $n$ and $p$, the respective equilibrium concentrations $n_0$ and $p_0$, which are connected to each other via $n = n_0 + \Delta n$ and $p = p_0 + \Delta n$ with the excess carrier concentration $\Delta n$, given by
\begin{equation}
    \Delta n = n_i\left(\exp\frac{eV}{2kT}-1\right).
\end{equation}
The effective intrinsic carrier concentration is given by 
\begin{equation}
    n_i = n_0 = n_p = n_{i,0}\exp\frac{\Delta E_\text{Si}}{2kT}
\end{equation}
with $n_{i,0} = 8.28\cdot10^9\,\text{cm}^{-3}$ and $\Delta E_\text{Si} = 0.005$\,eV, which accounts for \emph{bandgap narrowing} \cite{Richter2013ReassessmentCells}. Finally, the coefficient $B_\text{tot} = (1-P_\text{PR})B_\text{low}$ contains the radiative recombination coefficient for low-doped Si, $B_\text{low} = 4.73\cdot10^{-15}\,\text{cm}^3\text{s}^{-1}$,\cite{trupke:2003} and the \emph{photon-recycling} probability $P_{PR}$, as described by Richter \emph{et al.}\cite{Richter2013ReassessmentCells}

The short-circuit current density for the silicon cell is given by 
\begin{equation}
    J_\text{sc, Si} = e\int_{\lambda_\text{pero}}^{\lambda_\text{Si}} A(\lambda)\Phi_f(\lambda)\,\text{d}\lambda + e\int_0^{\lambda_\text{Si}} A(\lambda)\Phi_b(\lambda)\,\text{d}\lambda + J_\text{LC}
\end{equation}
with the photon fluxes $\Phi_f$  and $\Phi_b$ reaching the front and back sides of the silicon cell, respectively, and the current density $J_\text{LC}$ due to \emph{luminescent coupling}, which is treated in apendix \ref{sec:lc} below. The absorption $A(\lambda)$ is calculated according to the Tiedje-Yablonovitch limit,\cite{tiedje:1984} hence we assume perfect light trapping.
\begin{equation}
    A(\lambda) = \frac{\alpha_\text{Si}(\lambda)}{\alpha_\text{Si}(\lambda)+\left(4n_\text{Si}^2d_\text{Si}^{}\right)^{-1}}.
\end{equation}
Here, $\alpha_\text{Si}$ and $n_\text{Si}$ are the absorption coefficient and refractive index of silicon, respectively. Assuming a sharp absorption edge as for the perovskite top cell would not be a good assumption for silicon because of its indirect bandgap characteristic, which makes silicon weakly absorbing for a large wavelength range.

\section{Luminescent coupling}
\label{sec:lc}

Luminescent coupling can affect the performance of high quality multi-junction solar cells significantly. In general, luminescent coupling describes a process, where photons generated by radiative recombination in a high-bandgap subcell are absorbed in an adjacent subcell with a lower bandgap. Different methods were developed to model luminescent coupling, most are based on equivalent circuits for diodes.\cite{Jia2015Bias-dependenceCells}

Here, we derive a simple model for luminescent coupling in a tandem solar cell with idealized top and bottom cells. As mentioned above, the top cell is described by a one-diode equation with infinite parallel and zero series resistance,
\begin{equation}
    \label{eq:iv-top}
    J_\text{top} = J_\text{ph,top} - J_0 \left[\exp\left(\frac{qV}{k_bT}\right)-1\right] = J_\text{ph,top} - J_\text{rec,top},
\end{equation}
where $J_\text{ph,top}$ is the photocurrent density from external recombination, $J_0$ is the saturation current density, $V$ is the bias voltage of the top junction and $J_\text{rec,top}$ is the recombination current density. The recombination of the idealized top coll is exclusively radiative. Therefore, the additional current density in the bottom cell because of luminescent coupling $J_\text{LC}$ can be described as
\begin{equation}
    J_\text{LC} = \eta_\text{LC} J_\text{rec,top},
\end{equation}
 where $\eta_\text{LC}$ is the luminescent-coupling efficiency. In an idealized solar cell this efficiency only depends on the optical properties of the solar cell, because light generated in the top cell only can escape the perovskite layer through the top, where is it lost; into the bottom cell, where it can be absorbed; or it is absorbed in another layer of the solar cell, as discussed in appendix \ref{sec:opt} and also shown in Fig.\ \ref{fig:lc-tandem-optical}. In real solar cells, depending on the type of semiconductor and the material quality, $\eta_\text{LC}$ is often dominated by electrical effects, because usually only a fraction of the recombination occurs radiatively.\cite{Sogabe2013Experimentalcells, Lim2013LuminescenceMeasurement}

The overall electrical current density in the bottom cell including LC is given by
\begin{equation}
    \label{eq:iv-bot}
    J_\text{bot}(V_\text{bot}) = J_\text{ph,bot} - J_\text{rec,bot}(V_\text{bot}) + \eta_\text{LC} J_\text{rec,top}(V_\text{top})
\end{equation}
If the tandem cell is build as a monolithic two-terminal cell, the electrical current densities flowing through the top and bottom cells are identical,
\begin{equation}
    J_\text{bot} \equiv J_\text{top} \equiv J_\text{tandem}
\end{equation}
In order to model the $J$-$V$ characteristic of a two-terminal cell with luminescent coupling, both eqn (\ref{eq:iv-top}) and (\ref{eq:iv-bot}) need to be satisfied at the same time. To find a solution for the cell voltage, the $J$-$V$ characteristic needs to be calculated as a function of the current density $J_\text{tandem}$ . Therefore, the voltage-dependent recombination current density needs to be inverted. While this is easily done for a one-diode model with a cell obeying the Shockley-Queisser limit, it is not straight forwards for the characteristic of a silicon solar cell according to considerations of Richter \emph{et al.}\cite{Richter2013ReassessmentCells} We use a numerical approach by solving the forward function in the dark for a very fine grid of voltages and use a linear interpolation to approximate the inverse function. With the inverse function approximation we can calculate the power of the cell,
\begin{equation}
    \label{eq:power-num}
    \begin{aligned}
        J_\text{rec,top} &= J_\text{ph,top} - J_\text{tandem},\\
        J_\text{rec,bot} &= J_\text{ph,bot} + \eta_\text{LC} \cdot J_\text{rec,top} - J_\text{tandem},\\
        P_\text{tandem} &= J_\text{tandem} \cdot \left[V_\text{top}(J_\text{rec,top}) + V_\text{bot}(J_\text{rec,bot}, J_\text{rec,top})\right].
    \end{aligned}
\end{equation}

Because all components of eqn (\ref{eq:power-num}) only depend on the current density of the tandem cell and the photocurrent densities generated in the top and bottom cells, the power can be computed very fast on a regular grid. For every time step of radiation data in the TMY3 time series we first calculate the photocurrent densities in the top and bottom cells, $J_\text{ph, top}$ and $J_\text{ph, bot}$. Subsequently we calculate the cell voltages of the subcells for a grid between 0.1 and 50\,mA/cm$^2$ with a resolution of 0.1\,mA/cm$^2$. With the sum of the voltages and the corresponding current density, the output power density is found, where the maximum is taken as \emph{maximum power point} (MPP) power output.

For the case of four-terminal cells the output power density is calculated independently for each subcell. To account for luminescent coupling first the maximum power output of the top cell is evaluated. The recombination current at this working point $J_\text{rec,top}^\text{mpp}$ is then used to calculate the current contribution in the bottom cell from luminescent coupling. This contribution is in general very low, because the radiative recombination at the maximum power point is low.
\begin{equation}
    \label{eq:power-num2}
    \begin{aligned}
        P_\text{top} &= J_\text{top} \cdot V_\text{top}(J_\text{rec,top}),\\
        P_\text{bot} &= J_\text{bot} \cdot V_\text{bot}(J_\text{rec,bot}, J_\text{rec,top}^\text{mpp})\\
        P_\text{tandem} &= P_\text{top} + P_\text{bot}.
    \end{aligned}
\end{equation}

\section{Optical modeling of luminescent coupling}
\label{sec:opt}

We performed optical simulations to assess, (1) how much of the light generated in the perovskite layer can leave a single-junction perovskite cell, and (2) how much light reaches the silicon layer in a perovskite/silicon tandem solar cell. The optical simulations are performed with the \texttt{MATLAB}-based package \texttt{GenPro4}, which is based on the net-radiation method.\cite{santbergen:2017} This package mimics the coherence properties of light by allowing the user to decide, in which layers light behaves coherently and incoherently, respectively.

\begin{figure}
 \centering
 \includegraphics{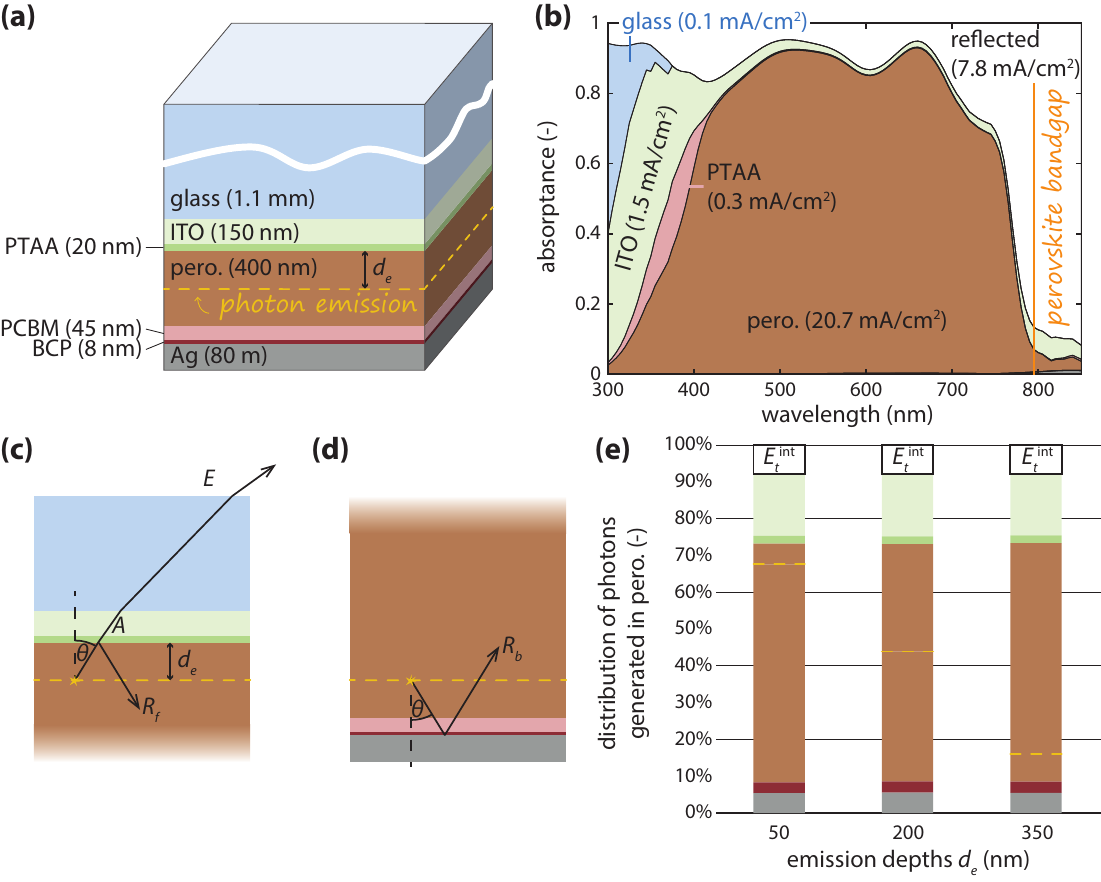}
 \caption{Optical simulations on single-junction perovskite solar cells. (a) Illustrating the solar cell layer stack, which is based on recent work by Liu \emph{et al.}\cite{liu_open-circuit_2019} (b) Absorption profile for this layer stack when light is incident from the air above the glass. The position of the perovskite bandgap (\ce{MAPbI3}, 1.56\,eV) is indicated. (c) Layer stack for assessing, how light emitted in the perovskite layer interacts with the layers above perovskite. The dashed yellow line marks the position of the light emission. (d) Same as (c), but for layers below the perovskite layer. (e) Relative distribution of photons with 795\, nm wavelength, which are isotropically emitted at three different positions in the perovskite layer. The emission depths has only little effect on the fraction of light reaching the silicon. The colors correspond to the legend in (a). The yellow dashed lines separate the fraction absorbed in the perovskite above and below the position of emission.}
\label{fig:lc-opt}
\end{figure}

Figure \ref{fig:lc-opt} illustrates the optical simulations for single-junction perovskite solar cells, based on the solar-cell structure by Liu \emph{et al.}\cite{liu_open-circuit_2019} The layer stack, shown in Fig.\  \ref{fig:lc-opt}(a), consists of Corning Eagle XG glass,\cite{cushman:2016} indium tin oxide (ITO), poly[bis(4-phenyl)(2,4,6-trimethylphenyl)amine (PTAA), methylammonium lead iodide (\ce{MAPbI3}) as perovskite,\cite{guerra:2017} phenyl-\ce{C61}-butyric acid methyl ester (PCBM), bathocuproine (BCP),\cite{liu:2005} and silver.\cite{johnson:1972} For ITO, PTAA and PCBM, we used in-house measured $n,k$ data. The glass substrate was treated incoherently, all other layers were treated incoherently by \texttt{GenPro4}. Figure \ref{fig:lc-opt}(b) shows the absorption of this structure, when light is incident via the glass side.

To calculate luminescence in the solar cell, we have to combine two sets of simulations. Figure \ref{fig:lc-opt}(c) shows the simulation setup, where we study, how light emitted into the upper hemisphere interacts with the layer stack on the front side of the solar cell. To simulate emission within the perovskite layer, we assume an infinitely thick perovskite layer on bottom, from which light is incident on the structure, because \texttt{GenPro4} cannot treat light emitted from within the structure. From this simulation, we can derive the fraction $E(\theta)$ of light, that leaves the solar cell when light has an angle of incidence $\theta$ in the perovskite layer. Further, the simulation delivers the reflection of the front layers $R_f(\theta)$ and the absorption $A(\theta)$ for all layers. Note that the simulation accounts for absorption in the perovskite above the position of emission. Figure \ref{fig:lc-opt}(d) shows the simulation setup for the lower layers of the solar cell for light that is emitted into the lower hemisphere. To mimic the emission within the perovskite, light is simulated as incident via an infinitely thick perovskite layer on top. From this simulation, we retrieve the reflection from the back $R_b(\theta)$ and -- if wanted the absorption in the different layers. 

To calculate the total amount of escaping light $E_t(\theta)$, we also must take reflection from the back side into account, leading to a geometric series,
\begin{subequations}
\label{eq:abs}
\begin{equation}
    \label{eq:abs_front}
    E_t = E\left(1+R_b\right)\left[1+R_fR_b+\left(R_fR_b\right)^2+\ldots\right]=\frac{E\left(1+R_b\right)}{1-R_fR_b},
\end{equation}
where we omitted the dependency on $\theta$ for brevity. The summand $R_b$ in the numerator accounts for light, which is emitted into the lower hemisphere and reflected back. Note that we can use eqn (\ref{eq:abs_front}) also for calculating the total absorption in the layers above the position of emission via replacing $E$ with the respective absorptance $A$. For calculating the total absorption in layers below the position of emission, we have to slightly adapt eqn (\ref{eq:abs_front}) and find
\begin{equation}
    \label{eq:abs_back}
    A_t =\frac{A\left(1+R_f\right)}{1-R_fR_b}.
\end{equation}
\end{subequations}
Note that we omit effects due to coherence in eqn (\ref{eq:abs}). Finally, we have to integrate over all angles, because light will be emitted isotropically in the perovskite layer.
\begin{equation}
    E_t^\text{int} = \frac{1}{4\pi}\int_0^{2\pi}\int_0^\frac{\pi}{2} E_t\,\sin\theta\,\d\theta\,\d\phi.
\end{equation}
By replacing $E_t$ with the total absorptance $A_t$ for a specific layer, the integrated absorption in that layer can be calculated. Note that the integral only extends over hemisphere, and not over the full sphere, because the other sphere is accounted for in the summands $R_b$ and $R_f$ in the nominators of eqn (\ref{eq:abs}).

Figure \ref{fig:lc-opt}(e) shows, where light emitted in the perovskite layer ends up for three different depths of emission $d_e$. We see that $d_e$ hardly affects the picture. For all depths, the emitted fraction is $E_t^\text{int} = 7.8\%$. Around 65\% are reabsorbed in the perovskite layer. The largest parasitic absorption occurs in the ITO, where around 17\% are lost. If no light was reflected from the back ($R_b = 0$) around 3.6\% of the emitted light would leave the layer stack. This fraction is less then half of $E_t^\text{int}$, because light, which is originally emitted into the upper hemisphere but then reflected back into the lower hemisphere, can be reflected upwards again when the back is reflected but it is lost otherwise. The low fraction of emission can be explained by total internal reflection. At 795\,nm wavelength, for which we study luminescence, the refractive index of the perovskite is $n=2.47$.\cite{guerra:2017} Hence, all light that is emitted into angles larger than $\theta>23.83$\textdegree\ is trapped inside they layer stack because of total internal reflection. Note, that we did not consider photon recycling here: light, that is reabsorbed in the perovskite can lead to another generated photon.

\begin{figure}
 \centering
 \includegraphics{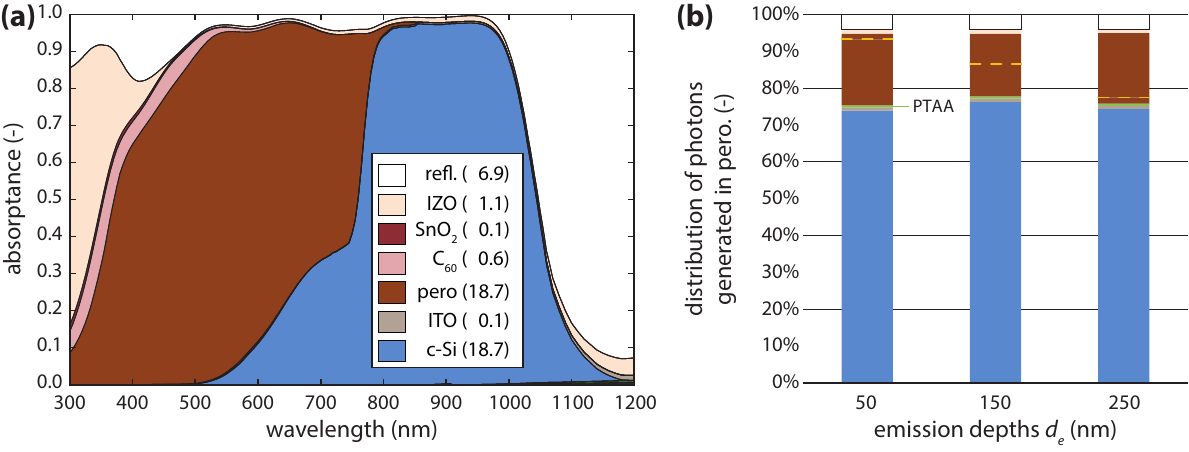}
 \caption{More details on the optical simulations on the tandem solar cell structure, which is discussed in Fig.\ \ref{fig:lc-tandem-optical}. (a) The absorption profile for the tandem solar cell structure shown in Fig.\ \ref{fig:lc-tandem-optical}(a), when light is incident from top. The numbers in brackets indicate the photocurrent density equivalent to the absorption in the layer in mA/cm$^2$. (b) Relative distribution of photons with 795\, nm wavelength, which are isotropically emitted at three different positions in the perovskite layer. The emission depths has only little effect on the fraction of light reaching the silicon. The colors correspond to the legend in (a) except PTAA, which is not visible in (a). The yellow dashed lines separate the fraction absorbed in the perovskite above and below the position of emission.} 
\label{fig:lc-tandem-optical-detail}
\end{figure}

The layer stack considered for the simulations of the tandem solar cell is shown in Fig.\ \ref{fig:lc-tandem-optical}(a): lithium fluoride (LiF),\cite{li:1976} indium zinc oxide (IZO), silicon oxide (\ce{SnO2}), \ce{C60}, \ce{MAPbI3},\cite{guerra:2017} PTAA, ITO, intrinsic amorphous hydrogenated silicon ($i$ a-Si:H), crystalline silicon (c-Si), \cite{green:1995} $i$ a-Si:H, $p$-doped a-Si:H, aluminium-doped zinc oxide (AZO), and silver (Ag).\cite{johnson:1972} For all materials, where no reference is given, the $nk$-data were determined in-house.   

Figure \ref{fig:lc-tandem-optical-detail} shows more details on the optical simulations on the tandem solar cell structure, which is discussed in Fig.\ \ref{fig:lc-tandem-optical} of the main manuscript. Figure \ref{fig:lc-tandem-optical-detail}(a) shows the absorption profile for the tandem solar cell structure shown in Fig.\ \ref{fig:lc-tandem-optical}(a), when light is incident from top. The layer thicknesses were adapted such, that we have current matching between top and bottom cells for front-side illumination under STC. Figure \ref{fig:lc-tandem-optical-detail}(b) shows, how the emission depth of the light in the perovskite layer affects the fractions of the emitted light ending up in the different layers. As for the single-junction cells [Fig.\ \ref{fig:lc-opt}(e)], the emission depths has only little effect.

\section*{Conflicts of interest}
There are no conflicts to declare.

\section*{Acknowledgements}
P. T. thanks the Helmholtz Einstein International Berlin Research School in Data Science (HEIBRiDS) for funding. We acknowledge the German Federal Ministry for Education and Research (BMBF) for support from the SNaPSHoTs project in the framework of the German-Israeli bilateral R\&D cooperation in the field of applied nanotechnology (grant no. 01IO1806). The results were obtained at the Berlin Joint Lab for Optical Simulations for Energy Research (BerOSE) and the Helmholtz Excellence Cluster SOLARMATH of Helmholtz-Zentrum Berlin f\"{u}r Materialien und Energie, Zuse Institute Berlin and Freie Universit\"{a}t Berlin.

\bibliographystyle{aipnum4-1}
\bibliography{local_bib} 

\end{document}